\documentclass[12pt]{iopart}

\usepackage{iopams} 
\usepackage{graphicx}
\usepackage{hyperref}

\def\Xint#1{\mathchoice
{\XXint\displaystyle\textstyle{#1}}%
{\XXint\textstyle\scriptstyle{#1}}%
{\XXint\scriptstyle\scriptscriptstyle{#1}}%
{\XXint\scriptscriptstyle\scriptscriptstyle{#1}}%
\!\int}
\def\XXint#1#2#3{{\setbox0=\hbox{$#1{#2#3}{\int}$}
\vcenter{\hbox{$#2#3$}}\kern-.5\wd0}}

\def\dashint{\Xint-}

\eqnobysec

\begin{document}

\title{A random wave model for the Aharonov-Bohm effect}

\author{Alexander J~H~Houston, Martin Gradhand and Mark R~Dennis}

\address{H H Wills Physics Laboratory, University of Bristol, Bristol BS8 1TL, UK}

\begin{abstract}
We study an ensemble of random waves subject to the Aharonov-Bohm effect.
The introduction of a point with a magnetic flux of arbitrary strength into a random wave ensemble gives a family of wavefunctions whose distribution of vortices (complex zeros) are responsible for the topological phase associated with the Aharonov-Bohm effect. 
Analytical expressions are found for the vortex number and topological charge densities as functions of distance from the flux point.
Comparison is made with the distribution of vortices in the isotropic random wave model.
The results indicate that as the flux approaches half-integer values, a vortex with the same sign as the fractional part of the flux is attracted to the flux point, merging with it at half-integer flux.
Other features of the Aharonov-Bohm vortex distribution are also explored.
\end{abstract}

%%% mrd: QM: topological and geometric phases, random/stochastic phenomena

\vspace{2pc}

%\submitto{\JPA}

\section{Introduction}\label{sec:int}

Random waves provide a universal model describing diverse stochastic phenomena. 
A particular case is the isotropic random wave model \cite{l-h:statistical,ww:acoustics,bd:isotropic}, representing an ensemble of solutions of the Helmholtz equation (time-independent wave equation) which is ergodic and statistically invariant to euclidean transformations, and whose only lengthscale is the wavenumber $k$. 
An isotropic random function contrasts with a plane wave, which propagates in a specified direction; this is similar to the difference between surface oscillations in the bulk of the ocean \cite{l-h:isotropic}, and waves breaking close to the shore. 
A major application of the random wave model follows from Berry's hypothesis \cite{berry:regular} that random waves quantitatively resemble high-energy modes and resonances of systems whose classical dynamics is chaotic, and much effort has gone into exploring this under various conditions \cite{ww:acoustics,stoeckmann,bs:autocorrelation,berry:statistics,ksw:classical,ur:supporting}. % more refs?
As such, the random wave model plays an analogous role in the quantum chaology of eigenfunctions to the random matrix hypothesis for eigenvalues, i.e.~that spectra of gaussian random matrices give a good model of the distributions of energy eigenvalues for quantum chaotic systems, orthogonal or unitary depending on whether time-reversal symmetry is preserved or broken \cite{bohigas:chaos}.

Here we explore an ensemble of random wavefunctions in which, like waves at the shore, time-reversal symmetry is broken: two-dimensional solutions of a charged, scalar quantum wave in the presence of a point of electromagnetic flux, that is, experiencing the Aharonov-Bohm (AB) effect \cite{AB,op:quantum}.
Elementary arguments show that on circling the flux point, a quantum particle should acquire a phase proportional to its charge times the flux strength. 
Being independent of the path geometry, the AB effect is an important instance of topology in physics, as well as being useful in the study of the structure of materials, as it causes their resistance to be oscillatory and flux-dependent \cite{aa:AB,bachtold:AB,peng:AB}. 
High-energy modes of chaotic billiards with AB flux lines are naturally complex, unlike usual real-valued cavity modes. 
They have been studied numerically in detail \cite{br:statistics,br:dislocations}, and Berry's hypothesis suggests the results of the complex AB random wave ensemble should agree with the behaviour of such systems.

The complex AB phase factor acquired by a quantum particle of charge $q$ experiencing a point flux $\Phi$ is $\exp(\rmi \alpha)$, depending on the dimensionless flux parameter $\alpha = q \Phi/\hbar$.
Around the flux, any quantum wavefunction ought to be single-valued and continuous, which appears inconsistent with the AB phase when $\alpha$ is not an integer \cite{AB,berry:exact}. 
The celebrated AB wavefunction \cite{AB} resolves this problem; it is a complex, single-valued solution of the Schr\"odinger equation for a plane wave scattered by the flux at the origin. 
When the quantum wave is initially travelling in direction $\Theta$, it is given in polar coordinates $r,\theta$ by
\begin{equation}
   \psi_{\mathrm{AB}}(r,\theta;\alpha,\Theta)=\sum_{\ell=-\infty}^{\infty}(-\rmi)^{|\ell-\alpha|}J_{|\ell-\alpha|}(kr)\exp(\rmi\ell[\theta+\pi-\Theta]),
   \label{eq:psiAB}
\end{equation}
where $J_{\nu}$ denotes a Bessel function of first kind of not necessarily integer order $\nu$.
For $kr \gg 1$, the wavefunction is $\psi_{\rm{AB}} \sim \exp(\rmi k r \cos(\theta-\Theta) + \rmi \alpha [\theta-\Theta]) + O(r^{-1/2})$ \cite{AB,berry:wavefront}, i.e.~a plane wave multiplied by the AB phase factor, so an integral of the phase $\chi = \arg \psi_{\mathrm{AB}}$ on a large loop around the origin gives $2\pi\alpha$.  
On the other hand, integrating $\chi$ on a small loop around the origin gives $2\pi$ times $n$, the nearest integer to $\alpha$ (i.e.~$|\alpha - n| \le \frac{1}{2}$).

\begin{figure}%[!ht]
\centering
\includegraphics[height = 280pt]{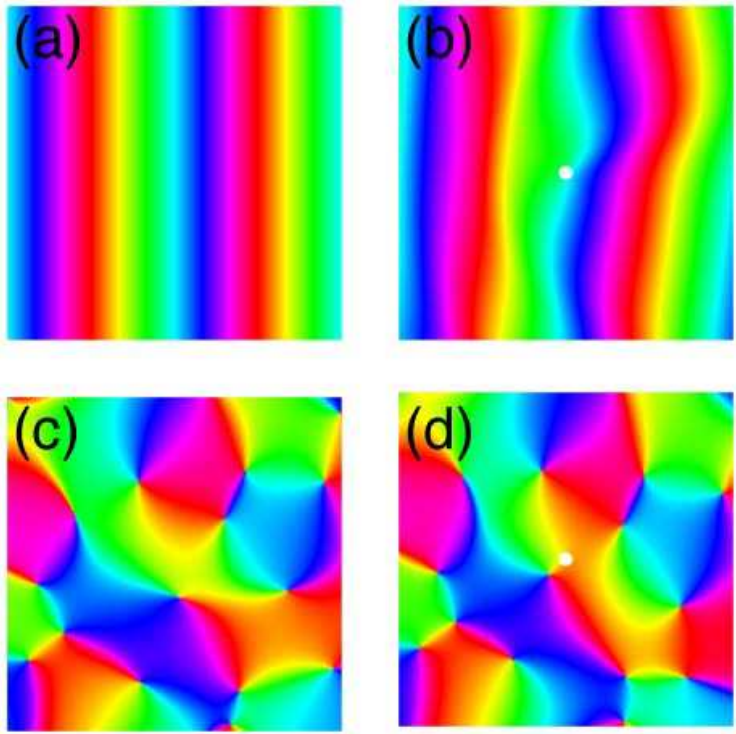} 
\caption{
   The phase $\chi = \arg\psi$ of deterministic and random waves, both free and scattered by a flux point.
   (a) Plane wave $\exp(\rmi k r \cos(\theta))$ travelling left to right; 
   (b) AB wavefunction (\ref{eq:psiAB}) for plane wave travelling left to right $(\Theta = 0)$; 
   (c) Isotropic random wave (\ref{eq:Psirw});
   (d) AB random wavefunction (\ref{eq:PsiAB}) with same random variables $a_j,\theta_j$.
   Phases are represented by hue, and in (b) and (d), the flux point, represented by a white disk, has $\alpha = 0.25$.
   Each square has side length $4\pi$ in units of $R = kr$.
   \label{fig:ABphase}}
\end{figure}

The regular phasefronts of a plane wave (Figure \ref{fig:ABphase} (a)) are disrupted on encountering a flux point (Figure \ref{fig:ABphase} (b)).
As $r$ increases, this distortion increasingly resembles a plane wave together with a noninteger phase discontinuity in the shadow $\theta = \Theta$, leading to a nonintegral AB phase factor in the limit $r \to \infty$. 
These deterministic phase patterns appear very different from those of random waves, shown in Figure \ref{fig:ABphase} (c) for the isotropic random model, and its AB analogue in Figure \ref{fig:ABphase} (d) (defined below in (\ref{eq:Psirw}) and (\ref{eq:PsiAB})).
Away from the flux, the two random wave patterns look qualitatively similar, so it would seem that the effect of the flux is rather subtle.
Given this resemblance, how does the AB phase manifest itself for random waves?
What is the quantitative difference between the isotropic random wave model and the AB random waves?
How much does the flux alter the statistics of eigenfunctions in chaotic AB billiards, other than making them complex?
It is these questions we address in the following.

Phase patterns of complex random waves are dominated by their \emph{vortices} (wave dislocations), i.e.~the nodal points where the phase $\chi$ is undefined \cite{bd:isotropic,br:dislocations}. 
On a right-handed circuit around each vortex point, the phase changes generically by $2\pi$ times a signed integer (statistically almost always $\pm1$). 
This integer is called the \emph{topological charge} (or strength) of the vortex; it is superficially similar to the AB phase, although the latter is determined by the external flux and may be nonintegral (whence the wavefunction is nondifferentiable at $0$).
The total topological charge of all the vortices in an area $\mathcal{A}$ is equal to the integral of the phase $\chi$ around the boundary $\partial \mathcal{A}$, 
\begin{equation}
   \oint_{\partial\mathcal{A}}\rmd\chi =\int_{\mathcal{A}}\left[\nabla\times\nabla\chi\right] \rmd^2\bi{r} = 2\pi \int_{\mathcal{A}}\left[\sum_i s_i\delta(\bi{r} -\bi{r}_i)\right]\rmd^2\bi{r}.
   \label{eq:chiint}
\end{equation}
following from from Stokes' theorem, since the phase gradient $\nabla\chi$ circulates with the sense of the topological charge and therefore has a finite nonzero curl when appropriately regularised \cite{mrd:burgers}.
The index $i$ labels the vortices in $\mathcal{A}$ with positions $\bi{r}_i$ and charges $s_i$.
The AB effect suggests this integral is non-zero whenever $\partial\mathcal{A}$ encloses a flux point, so there must also be a topological charge there.
The form of the deterministic AB wavefunction (\ref{eq:psiAB}) implies \cite{berry:wavefront} that the topological charge of the flux is the nearest integer $n$ to the flux strength $\alpha$, and therefore changes discretely at half-integer values. 
Similarly, the location and strength of the phase vortices (i.e.~the wavefront topology), is gauge-independent although the geometry of the phase pattern is not, since it is altered by changing the phase gradient.
The observable content of the phase is therefore contained in the vortex structure and this ought to be all that is needed for a complete description of the Aharonov-Bohm effect in a random wave.

Measures of the statistical vortex distribution are the average \emph{topological charge density} $\rho(\bi{r})$ and \emph{vortex density} $\Delta(\bi{r})$ \cite{bd:isotropic,fw:critical,sbs:distribution} which are smooth functions, % more refs?
\begin{equation}
   \rho(\bi{r}) = \left\langle\sum_i s_i\delta(\bi{r}-\bi{r}_i)\right\rangle, \qquad \Delta(\bi{r}) = \left\langle\sum_i \delta(\bi{r}-\bi{r}_i)\right\rangle,
   \label{eq:rhoDeltadef}
\end{equation}
where $\langle \bullet \rangle$ denotes ensemble averaging.
Random AB waves are statistically invariant to rotation about the flux at $0$, so $\rho$ and $\Delta$ only depend on radius $r$---or, more conveniently, $R = kr$---and $\alpha$; we therefore write $\rho(R,\alpha)$, $\Delta(R,\alpha)$.
While the topological charge at the flux point (or anywhere else) can never be fractional, we would expect to see the effect of non-integer $\alpha$ (at least for $R \gg 1$) in the statistical behaviour of the vortex structure, captured by $\rho$.
We choose to exclude the topological charge of $n$ at the origin in our definition of $\rho$ and $\Delta$; only standard phase vortices contribute to (\ref{eq:rhoDeltadef}).
The area integral of $\rho$ over the centred disk of radius $R$ (excluding the origin) should be equal, by (\ref{eq:chiint}), to the mean integral of phase $\chi$ on a circle of radius $R$ minus $2\pi n$, and therefore the vortex distribution ought to account for the fractional part of the AB phase.

As we have excluded the charge at the origin from the definition, $\rho$ and $\Delta$ depend on the fractional part of $\alpha$ (i.e.~they are periodic in $\alpha$ with period $1$), like all physical observables, but unlike the (unobservable) phase gradient which depends on $\alpha$ itself.
It will therefore be useful to write $\alpha = n + \beta$, for $-\frac{1}{2} < \beta \le \frac{1}{2}$.
Thus $\psi_{\rm{AB}}(r,\theta;\alpha,\Theta) = \exp(\rmi n [\theta+\pi-\Theta]) \psi_{\rm{AB}}(r,\theta;\beta,\Theta)$; in particular, if $\alpha = n \neq 0$, the wave is simply the incident plane wave times the integer AB phase factor.

The isotropic random wave ensemble can be defined, by analogy with Fourier's theorem, as sums of very many plane waves with random directions and amplitudes,
\begin{equation}
   \Psi_{\rm{irw}}(\bi{r}) = \frac{1}{\sqrt{N}}\sum_{j=1}^{N}a_j \exp(\rmi\bi{k}_j \cdot\bi{r}),
   \label{eq:Psirw}
\end{equation}
where $N \gg 1$, $\bi{k}_j$ are wavevectors of fixed magnitude $k$ with uniformly random directions $\theta_j$, and $a_j$ are independent identically distributed complex random variables, chosen to be circular gaussian random with $\langle a_j a_{j'}^* \rangle = \delta_{j j'}$, so $\langle |\Psi_{\rm{irw}}(\bi{r})|^2 \rangle = 1$.
When $N \gg 1$, by the central limit theorem $\Psi_{\rm{irw}}(\bi{r})$ is a complex gaussian random variable at each $\bi{r}$, so averages can often be calculated analytically.
The statistical distribution of the vortices in this system has been studied in depth \cite{bd:isotropic,br:dislocations,fw:critical,sbs:distribution,mrd:thesis,foltin:signed,mrd:correlations,mrd:nodal}; the isotropic vortex density $\Delta_{\rm{irw}} = 1/4\pi$ (in dimensionless units) and charge density $\rho_{\rm{irw}} = 0$ are obviously independent of position; 2-point vortex correlation functions have a richer structure, and will be discussed below.

Random AB wavefunctions are defined like (\ref{eq:Psirw}), but now are sums of AB waves (\ref{eq:psiAB}), with complex gaussian amplitudes $a_j$ and uniformly random directions $\theta_j$, 
\begin{equation}
   \Psi_{\rm{AB}}(\bi{r};\alpha) 
   = \frac{1}{\sqrt{N}}\sum_{j=1}^{N}a_j\psi_{\rm{AB}}(r,\theta;\alpha,\theta_j) 
   = \sum_{\ell} c_{\ell} J_{|\ell-\alpha|}(kr)\exp(\rmi \ell \theta), \label{eq:PsiAB} 
\end{equation}
where here and hereafter $\ell$ sums from $-\infty$ to $\infty$ unless otherwise stated, and the coefficients $c_{\ell} = (-\rmi)^{|\ell-\alpha|} N^{-1/2} \sum_{j = 1}^N a_j \exp(-\rmi \ell \theta_j)$ are complex gaussian random variables satisfying $\langle c_{\ell} c^*_{\ell'}\rangle = \delta_{\ell\ell'}$ if $\langle a_j a_{j'}^*\rangle = \delta_{jj'}$.
$\Psi_{\rm{AB}}(\bi{r};\alpha)$ indeed has the various properties previously discussed, and just like (\ref{eq:psiAB}), $\Psi_{\rm{AB}}$ has a non\-differentiable singularity at $r = 0$ when $\alpha \neq n$.
Both the deterministic and random AB wavefunctions have integer topological charge $n$ at the origin so the distribution of $\rho$ must therefore account for the fractional part of the AB phase as $R \to \infty$.
When the flux is zero, $\Psi_{\rm{AB}}(\bi{r};0) = \Psi_{\rm{irw}}(\bi{r})$, albeit expanded in a cylindrical basis; by Graf's addition theorem, any point of an isotropic random wave can be the origin of such an expansion \cite{msg:avoided}.

We will compare the random AB wavefunctions (\ref{eq:PsiAB}) with their isotropic counterparts (\ref{eq:Psirw}) using $\rho(R,\beta), \Delta(R,\beta)$ to recover the AB effect.
Figure \ref{fig:ABphase} (d) represents the random wave with the same random set of incident plane waves as the isotropic random wave in Figure \ref{fig:ABphase} (c), but scattered by a flux of $\alpha = \frac{1}{4}$. 
Vortices close to the flux point apparently are shifted, with the more distant vortices tending to be less affected.
The $+1$ vortex closest to the flux experiences the largest shift, \emph{towards} the origin.
We will explore this effect quantitatively by calculating $\rho(R,\beta)$ and $\Delta(R,\beta)$, and comparing with the isotropic model centred at a vortex.
This naturally leads to a discussion of the behaviour of the AB random waves for small $R$ in Section \ref{sec:near}, before we show how the Aharonov-Bohm effect is recovered in Section \ref{sec:recovering}, concluding with a discussion and outlook.

\section{Phase gradient and vortex statistics}\label{sec:statistics}

The topological charge density $\rho(R,\beta)$ and vortex density $\Delta(R,\beta)$, as functions of $R = k r$ and the fractional part of the flux $\beta$ can be found using the ensemble averages of the real and imaginary parts of $\Psi_{\mathrm{AB}} = \xi + \rmi \eta$, as described in detail in previous studies (e.g. \cite{bd:isotropic,mrd:nodal}).
They are
\begin{eqnarray}
   \rho(R,\beta) & = & \left\langle \delta(\xi)\delta(\eta) R^{-1}(\partial_R \xi \partial_{\theta} \eta - \partial_{\theta}\xi\partial_R \eta)\right\rangle,\label{eq:rhoxieta} \\
   \Delta(R,\beta) & = & \left\langle \delta(\xi)\delta(\eta) R^{-1}\left|\partial_R \xi \partial_{\theta} \eta - \partial_{\theta}\xi\partial_R \eta\right|\right\rangle. \label{eq:Deltaxieta}
\end{eqnarray}
The calculations therefore involve $\alpha$- and $R$-dependent variances of the fields and their derivatives, and, by ensemble averaging of (\ref{eq:PsiAB}), the relevant nonvanishing ones are
\begin{eqnarray}
   a = & \left\langle \xi^2 \right\rangle = \left\langle \eta^2 \right\rangle 
   & = \frac{1}{2}\sum_{\ell} J^2_{|\ell-\alpha|}(R), \label{eq:adef} \\
   b = & \left\langle \xi \partial_R \xi \right\rangle = \left\langle \eta \partial_R \eta \right\rangle 
   & = \frac{1}{2}\sum_{\ell}J_{|\ell-\alpha|}(R) J'_{|\ell-\alpha|}(R), \label{eq:bdef} \\
   c = & \left\langle \xi R^{-1}\partial_{\theta} \eta \right\rangle = -\left\langle \eta R^{-1}\partial_{\theta} \xi \right\rangle 
   & = \frac{1}{2R}\sum_{\ell}\ell J^2_{|\ell-\alpha|}(R), \label{eq:cdef} \\
   d = & \left\langle (\partial_R \xi)^2 \right\rangle = \left\langle (\partial_R \eta)^2 \right\rangle & = \frac{1}{2}\sum_{\ell} J'^2_{|\ell-\alpha|}(R), \label{eq:ddef} \\
   e = & \left\langle \partial_R\xi R^{-1}\partial_{\theta} \eta \right\rangle = -\left\langle \partial\eta R^{-1}\partial_{\theta} \xi \right\rangle 
   & =  \frac{1}{2R}\sum_{\ell} \ell J_{|\ell-\alpha|}(R)J'_{|\ell-\alpha|}(R), \label{eq:edef} \\
   f = & \left\langle R^{-2}(\partial_{\theta}\xi)^2 \right\rangle = \left\langle R^{-2}(\partial_{\theta}\eta)^2 \right\rangle 
   & = \frac{1}{2R^2}\sum_{\ell}\ell^2 J^2_{|\ell-\alpha|}(R). \label{eq:fdef}
\end{eqnarray} 
In the case of integer flux $\alpha = n$ (i.e.~$\beta = 0$), statistics are equivalent to the isotropic random wave model, and the sums have simple form: $a = \frac{1}{2}$, $d = f = \frac{1}{4}$, $b = c = e = 0$.
In the general case, they can be calculated in closed form as expressions of sums and products of Bessel functions, together with a certain $_3F_2$ hypergeometric function, as given in \ref{sec:appa}, (\ref{eq:afull})--(\ref{eq:ffull}).

Although we could calculate $\rho$ from (\ref{eq:rhoxieta}), it is more direct to calculate the average of the curl of the phase gradient, then proceed to relate this to the phase change around a region by (\ref{eq:chiint}).
Explicit calculations of the average azimuthal and radial components of the phase gradient vector $\overline{\chi_{\theta}}, \overline{\chi_R}$ involve the quantities (\ref{eq:adef})-(\ref{eq:cdef}), calculated in \ref{sec:appa}, with the result that
\begin{equation}
\fl   \overline{\chi_{\theta}}(R,\alpha) = R^{-1}\langle\partial_{\theta}\chi\rangle = \frac{c}{a} = \frac{1}{R}\frac{\sum_{\ell}\ell J^2_{|\ell-\alpha|}(R)}{\sum_{\ell} J^2_{|\ell-\alpha|}(R)}, \qquad
   \overline{\chi_R}(R,\alpha)  = \langle\partial_R\chi \rangle = 0,  \label{eq:achi}
\end{equation}
so the ensemble averaged phase gradient is a pure circulation around the origin, as it would be for a single vortex there.
The charge density $\rho$ now follows from (\ref{eq:chiint}),
\begin{eqnarray}
   \rho(R,\alpha) & = & \frac{1}{2\pi}\left\langle \nabla\times\nabla\chi \right\rangle  = \frac{1}{2\pi R}\frac{\partial}{\partial R}\left\langle \partial_{\theta}\chi\right\rangle, \label{eq:rhocurl} \\
   & = & \frac{1}{\pi}\frac{a e - b c}{a^2}, \label{eq:rhoabce}
\end{eqnarray}
where in (\ref{eq:rhocurl}) the curl operator commutes with ensemble averaging, and (\ref{eq:rhoabce}) follows from (\ref{eq:achi}).
Of course, $\rho(R,\alpha = n) = 0$, as for the isotropic random wave model.
$\rho$ given here is dimensionless; it acquires a factor of $k^2$ as a density in terms of $r$.

The calculation of $\Delta$ in (\ref{eq:Deltaxieta}) is more involved due to the presence of the modulus sign.
Details are again provided in \ref{sec:appa}, and the result is
\begin{equation}
   \Delta(r,\alpha) = \frac{1}{2\pi}\frac{(bc-ae)^2+(ad-b^2)(af-c^2)}{a^2\sqrt{(ad-b^2)(af-c^2)}}. 
   \label{eq:Deltabcdef}
\end{equation}
Again, the integer flux case emulates the isotropic random wave, $\Delta(R,n) = \Delta_{\rm{irw}} = 1/4\pi$.

\begin{figure}%[htb!]
\centering
\includegraphics[height=120 pt]{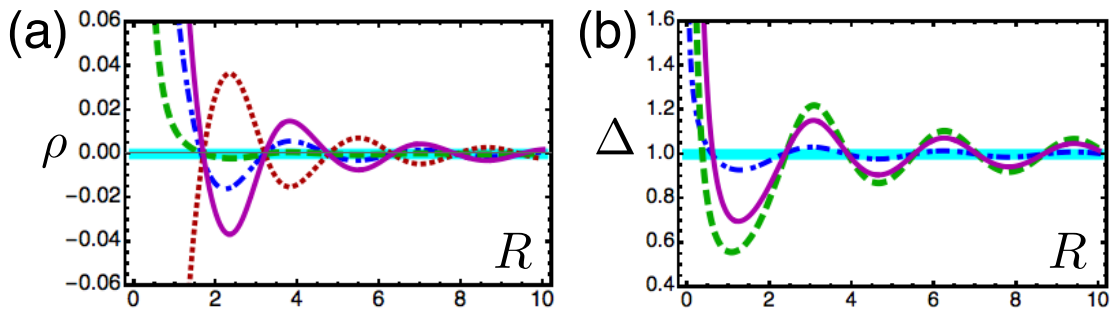}
\caption{
   Statistical vortex distributions, normalized with respect to the isotropic random wave density $\Delta_{\rm{irw}}$.
   (a) Radial topological charge density $\rho(R,\alpha)$; (b) radial vortex density $\Delta(R,\alpha)$.
   Values of fractional part of the flux $\beta$ are $0.25$ (solid line), $-0.25$ (dotted), $0.49$ (dashed), $0.05$ (dot-dashed) and $0$ (thick).
   \label{fig:rhoDelta}}
\end{figure}

The normalized radial densities $\rho(R,\beta)/\Delta_{\rm{irw}}$ and $\Delta(R,\beta)/\Delta_{\rm{irw}}$ are plotted in Figure \ref{fig:rhoDelta} for several different choices of $\beta$.
In both cases, the densities tend to those of the isotropic ensemble results as $R \to \infty$ ($\rho \to 0, \Delta \to \Delta_{\rm{irw}}$), since far from the influence of the flux $\beta$ each scattered AB wavefunction is almost a plane wave.
Features of the dependence of $\rho$ and $\Delta$ on $\beta$ can be understood from first principles: when $\beta = 0$ ($\alpha = n$), there is no time-reversal symmetry breaking, so the distributions of positive and negative vortices are equivalent and $\rho(R,n) = 0$.
Furthermore, sending $\beta \to -\beta$ (equivalent to a time-reversal operation) reverses vortex charges, so we expect 
\begin{equation}
   \rho(R,-\beta) = -\rho(R,\beta) \qquad \hbox{and} \qquad \Delta(R,\beta) = \Delta(R,-\beta).
   \label{eq:rDpm}
\end{equation}
These follow from considering the sums in (\ref{eq:adef})--(\ref{eq:fdef}) appearing in (\ref{eq:rhoabce}) and (\ref{eq:Deltabcdef}); reversing the sign of $\beta$ is equivalent to $\ell \leftrightarrow -\ell$ in each of the sums, so the signs of $c$ and $e$ reverse but those of $a,b,d,f$ do not.

It might be expected that $\beta=\frac{1}{2}$ would produce the largest fluctuations of $\rho$ since this is furthest from integer $\alpha$; the curve for $\beta=0.49$ in Figure \ref{fig:rhoDelta} (a) suggests this is not the case. 
Indeed, (\ref{eq:rDpm}) in combination with the $\alpha$ periodicity of $\rho$ gives $\rho(R,\frac{1}{2}) = 0$, since $\rho(R,\frac{1}{2}) = \rho(R,-\frac{1}{2}) = -\rho(R,\frac{1}{2})$. 
The `false time-reversal symmetry breaking' of $\beta = \pm\frac{1}{2}$ unbiases the sign of $\rho$ everywhere, like an isotropic random wave, though the random wave model remains complex-valued.
This differs from eigenfunctions of chaotic AB eigenfunctions with standard boundary conditions, which can be written as real functions times an overall phase factor, with corresponding gaussian orthogonal ensemble eigenvalue statistics \cite{rb:false}. 
There is no analogous effect on $\Delta(R,\beta=\frac{1}{2})$, and as may be inferred from Figure \ref{fig:rhoDelta} (b), $\beta = \frac{1}{2}$ has the largest fluctuations in vortex density.

Asymptotic expressions for $\rho$ and $\Delta$ may be found using the analytic forms of $a, \ldots, f$, for which we assume $\beta \ge 0$ without loss of generality.
For large values of $R$,
\begin{equation}
  \fl \rho \sim \frac{(1-2\beta)\sin(\beta \pi)}{2 \pi^2} \frac{\sin(2R)}{R^2}, \qquad
   \Delta \sim \frac{1}{4\pi} + \frac{\sin(\beta \pi)}{2\pi^2} \frac{\cos(2R)}{R}, \qquad R \gtrsim 2\pi,
   \label{eq:largeR}
\end{equation}
tending to the isotropic random wave values as $R \to \infty$ and when $\beta = 0$.
As we have seen, when $\beta = \frac{1}{2}$, $\rho$ vanishes whereas variation around the asymptotic value of $\Delta$ is largest.
Furthermore, $\rho$ decays with $R^{-2}$, more rapidly than the $R^{-1}$ decay of $\Delta$, and their oscillations are out of phase.
For small $R$, the first terms of the expansions agree, 
\begin{equation}
   \rho, \, \Delta \approx  \frac{(1-2\beta) \Gamma^2(1+\beta)}{2^{2-4\beta}\pi\Gamma^2(2-\beta)}R^{-4\beta} 
   -  \frac{(1-2\beta)^2 \Gamma^4(1+\beta)}{2^{3-8\beta}\pi \Gamma^4(2-\beta)} R^{2-8\beta}. \label{eq:smallR} \\
\end{equation}
The first term diverges at $R = 0$ when $\beta > 0$ (the second when $\beta > \frac{1}{4}$); in fact, following terms (equal for $\rho,\Delta$) behave as $R^{2n-4(n+1)\beta}$, more of which diverge as $\beta$ approaches $\frac{1}{2}$; the coefficients all vanish at $\beta = \frac{1}{2}$.

\begin{figure}%[!htbp]
\centering
\includegraphics[height=120 pt]{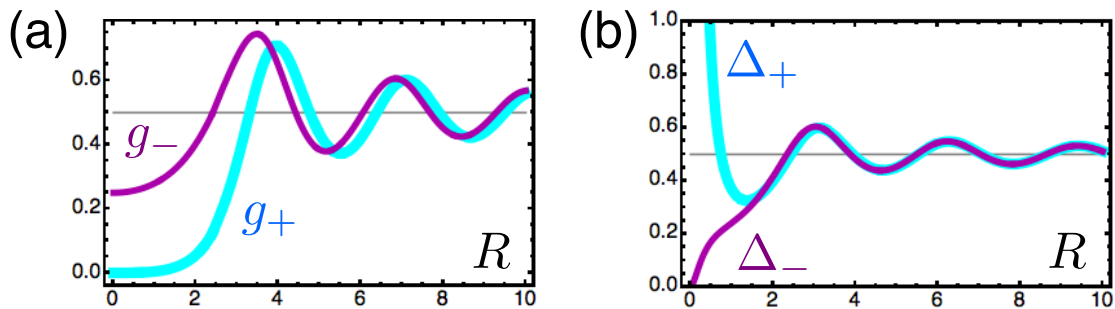}
\caption{
   Radial densities of positive and negative topologically charged vortices, assuming a topological charge at the origin.
   The thick lines (thin lines) represents the positive (negative) vortex density.
   (a) Isotropic random wave assuming a positive vortex at the origin.
   (b) AB random wave with $\beta = 0.375$, with densities normalized with respect to $\Delta_{\mathrm{irw}}$. 
   All densities approach $\frac{1}{2}$ as $R \to \infty$.
   \label{fig:±densities}}
\end{figure}

It is natural to compare these AB vortex densities with the corresponding densities of vortices in the isotropic random wave model at a distance $R$ from a vortex of fixed sign, say $+1$.
These densities are given by the 2-point correlation functions, with and without modulus signs (i.e.~topological charge signs) \cite{bd:isotropic}
\begin{eqnarray}
\fl   g_s(R) & = & \frac{1}{\Delta_{\mathrm{irw}}^2} \left\langle \delta(\xi_0)\delta(\eta_0) \left(\partial_x\xi_0 \partial_y\eta_0 - \partial_x\eta_0 \partial_y\xi_0 \right) \delta(\xi_1)\delta(\eta_1) \left(\partial_x\xi_1 \partial_y\eta_1 - \partial_x\eta_1 \partial_y\xi_1 \right) \right\rangle, \label{eq:gsdef} \\
\fl   g(R) & = & \frac{1}{\Delta_{\mathrm{irw}}^2} \left\langle \delta(\xi_0)\delta(\eta_0) \left|\partial_x\xi_0 \partial_y\eta_0 - \partial_x\eta_0 \partial_y\xi_0 \right| \delta(\xi_1)\delta(\eta_1) \left|\partial_x\xi_1 \partial_y\eta_1 - \partial_x\eta_1 \partial_y\xi_1 \right| \right\rangle, \label{eq:gdef}
\end{eqnarray}
where subscripts $0$ and $1$ denote quantities evaluated at $\bi{r}_0$ (fixed) and $\bi{r}_1$, and $R = k|\bi{r}_1-\bi{r}_0|$.
These functions are normalized with respect to $\Delta_{\mathrm{irw}}$ so, for instance, the number correlation $g\to_{R\to\infty} 1$.
Therefore $g_s(R)$ is analogous to $\rho(R)/\Delta_{\mathrm{irw}}$, giving the average topological charge density at $R$ from the fixed $+1$ vortex, and likewise $g(R)$ to $\Delta/\Delta_{\mathrm{irw}}$.
$g(R)$ can be expressed as an integral representation \cite{bd:isotropic} or in closed form as a complicated expression involving elliptic integrals \cite{mrd:thesis}.
The charge correlation function $g_s(R)$ is more straightforward \cite{bd:isotropic,fw:critical} and is $g_s(R) = 4 R^{-1} \partial\left[J_1^2(R)/(1 - J_0^2)\right]/\partial R$.
For large $R$, $g(R) \sim 1 + 4 \sin(2R)/\pi R$ \cite{hohmann:density}, contrasting with $\Delta(R,\beta)/\Delta_{\mathrm{irw}} \sim 1 + 2 \sin(\beta\pi) \cos(2R)/\pi R$; in addition to the $\beta$-dependence, the peaks of maximum and minimum density are shifted by $\pi/4$ in $R$.
Similarly, the asymptotic form $g_s(R) \sim 8\cos(2R)/\pi R^2$, has a similar relationship with $\rho(R,\beta)/\Delta_{\mathrm{irw}} \sim 2(1-2\beta)\sin(\beta \pi)\sin(2R)/\pi R^2$; in both cases, the charge oscillations decay more rapidly than the number oscillations, and are out of phase with them.
Both $g(R)$ and $g_s(R)$ are finite and nonzero at the origin, such that $g(R) + g_s(R) \approx O(R^2)$ \cite{bd:isotropic,mrd:thesis}.
This is because, statistically, vortices of like charge \emph{repel}, as can be seen from the like-charge correlation function $g_+(R) = \frac{1}{2}[g(R)+g_s(R)]$, with unlike correlation function $g_{-}(R) = \frac{1}{2}[g(R)-g_s(R)]$.
$g_+$ and $g_-$ are plotted in Figure \ref{fig:±densities} (b): unlike charges are more likely to surround the vortex for $R \lesssim \frac{\pi}{2}$, before both oscillate in phase, decaying as $R^{-1}$.

The analogous quantities for the AB random waves are the normalized positive and negative vortex densities $\Delta_{\pm}(R,\beta) = (2\Delta_{\mathrm{irw}})^{-1}[\Delta(R,\beta) \pm \rho(R,\beta)]$, plotted in Figure \ref{fig:±densities} (b) for $\beta = \frac{3}{8}$.
Assuming $0 < \beta \le \frac{1}{2}$, from the previous discussion we see that for large $R$ these oscillate in phase; for small $R$, $\Delta_+$ diverges as $R^{-4\beta}$ whereas $\Delta_- \approx O(R^{4\beta})$, corresponding to the earliest terms in the small-$R$ expansions of $\rho$ and $\Delta$ which do not agree.
This strong divergence of $\Delta_+$ suggests that a positive fractional flux attracts a positive vortex increasingly as $\beta$ approaches $\frac{1}{2}$; we have already discussed how the integer topological charge on the flux line increases by $+1$ when $\beta$ reaches $\frac{1}{2}$.

\section{Topological charge near the flux}\label{sec:near}

The divergence of the vortex density $\Delta(R,\beta)$ at $R = 0$ suggests a fractional flux attracts a vortex with a sign equal to $\mathrm{sign}\beta$. 
We have also discussed how the flux point has an integer topological charge $n$ which jumps by $\pm1$ when $\beta\to \pm\frac{1}{2}$.
In this section, we find it more convenient to define $\beta$ for $0 \le \beta \le 1$ rather than $-\frac{1}{2} \le \beta \le \frac{1}{2}$, so the topological charge of the flux depends whether $\beta < \frac{1}{2}$ (where it is $n$) or $\beta > \frac{1}{2}$ (where it is $n+1$).

We investigate the topological charge and vortex behaviour near the flux using a small-$R$ expansion of $\Psi_{\rm{AB}}(r,\theta;\alpha)$ in (\ref{eq:PsiAB}), noting that, from the limiting form of $J_{\nu}(R)$, the lowest order approximation is   
\begin{equation}
\fl   \Psi_{\rm{AB}}(R,\theta;\alpha) \approx \left(\frac{R}{2\rmi}\right)^{\beta} \exp(\rmi n\theta)\left[\frac{X_0}{\Gamma(1+\beta)}  + \frac{X_{+1}}{\Gamma(2-\beta)}\left(\frac{R}{2\rmi}\right)^{1-2\beta}\exp(\rmi \theta) \right],
   \label{eq:Psiapprox}
\end{equation}
where for integer $\ell$, $X_{\ell}$ are the complex random variables  
\begin{equation}
   X_{\ell} = \frac{1}{\sqrt{N}}\sum_{j=1}^N a_j \exp(\rmi[\gamma_j-(n+\ell)\theta_j]),
   \label{eq:XYdef}
\end{equation}
which are independent and identically distributed.
The topological charge at the origin is $n$ or $n+1$ depending on the relative magnitude of the two summands in $[ \bullet ]$ in (\ref{eq:Psiapprox}) close to the origin.
When $0<\beta<\frac{1}{2}$, the $\ell = 0$ term dominates so the flux charge is $n$; when $\beta >\frac{1}{2}$, the $+1$ term dominates so the charge is $n+1$.
If $\beta=\frac{1}{2}$, it is $n+1$ if $|X_{+1}|>|X_0|$, and $n$ otherwise; since their distributions are equivalent, averaging yields $n+\frac{1}{2}$. 
This confirms the previous statement that the number of vortices at the origin can only change when $\alpha=n+\frac{1}{2}$, in a way identical to that in the deterministic case \cite{berry:wavefront}. 
When $\beta > \frac{1}{2}$ it is natural to rearrange (\ref{eq:Psiapprox}) so the $\ell = +1$ term becomes the `constant' term in $[ \bullet ]$ in (\ref{eq:Psiapprox}), and the $\ell = -1$ term acquires a $\exp(-\rmi \theta)$ factor.

We wish to use (\ref{eq:Psiapprox}) to estimate the position of the nearest vortex to the origin.
An expansion of the form $A + B (x+\rmi y) + C (x - \rmi y)$ tends to have a $+1$ vortex close to the origin if $|B|$ is significantly larger than $|A|$ and $|C|$, and a $-1$ vortex if $|C|$ is larger; if $B$ and $C$ have equivalent magnitudes it is harder to predict.
When $\beta$ is near zero, however, the first term omitted from the expansion, proportional to $X_{-1}(R/2\rmi) \exp(-\rmi \theta)$ in $[ \bullet ]$ in (\ref{eq:Psiapprox}), competes with the $\ell = +1$ term, consistent with the random wave being close to an isotropic random wave whose topological charges are balanced.
However, as $\beta$ approaches $\frac{1}{2}$, this neglected term loses significance near the origin, solving the right hand side of (\ref{eq:Psiapprox}) is a good approximation of the nearest vortex, which is positive.

With this assumption, if the nearest vortex has coordinates $R_{\rm{nv}}$ and $\theta_{\rm{nv}}$, the angle $\theta_{\rm{nv}}$ is uniformly random, and depends on $\arg(X_0/X_{+1})$.
The radial distance $R_{\rm{nv}}$ satisfies
\begin{equation}
\frac{\Gamma(1+\beta)}{\Gamma(2-\beta)}\left(\frac{R_{\rm{nv}}}{2}\right)^{1-2\beta}=\frac{|X_0|}{|X_{+1}|}=\gamma,
\end{equation}
where the random variable $\gamma$ has a probability distribution function $P(\gamma)=2\gamma/(1+\gamma^2)^2$, since $|X_0|$ and $|X_{+1}|$ are independent Rayleigh distributed random variables. 
Making the appropriate change of variables, we find $R_{\rm{nv}}$ has the probability density
\begin{equation}
   P_{\beta}(R_{\rm{nv}}) = \frac{(1-2\beta)\Gamma^2(1+\beta)\Gamma^2(2-\beta)\left(R_{\rm{nv}}/2\right)^{1-4\beta}}{\left(\Gamma^2(2-\beta)+\Gamma^2(1+\beta)(R_{\rm{nv}}/2)^{2-4\beta}\right)^2}.
   \label{eq:PRnv}
\end{equation}
Of course, this is only valid when $0<\frac{1}{2} - \beta$ is small, and in fact the expectation value of $R_{\rm{nv}}$ with respect to the distribution (\ref{eq:PRnv}) diverges for $\beta > \frac{1}{4}$ (due to cases where $|X_0|$ is sufficiently larger than $|X_{+1}|$ that solving $\Psi_{\rm{AB}} = 0$ requires more terms in the expansion in $R$).
In this approximation, the approach of the nearest vortex to the flux as $\beta$ approaches $\frac{1}{2}$ can be captured by the conditional mean $\langle R_{\rm{nv}} \rangle_{\delta}$ of $R_{\rm{nv}}$ with respect to (\ref{eq:PRnv}) subject to $0 < R_{\rm{nv}}< \delta$, which is
\begin{equation}
\fl   \langle R_{\rm{nv}} \rangle_{\delta} =  z \delta \left( \,_2F_1\left(\begin{array}{c} 1 \quad 1 \\  1+\frac{1}{4\varepsilon} \end{array}; \frac{1}{1+ z} \right) - 1\right), \qquad \hbox{where } z = \frac{16^{\varepsilon} \Gamma^2(\frac{3}{2} + \varepsilon)}{\delta^{4\varepsilon} \Gamma^2(\frac{3}{2} - \varepsilon)},
   \label{eq:<Rnv>}
\end{equation}
with $\varepsilon = \frac{1}{2} - \beta$, and $_2F_1$ denotes a usual hypergeometric function.
Approximating this for small $\varepsilon$ using \cite{dlmf} \S15.12, $\langle R_{\rm{nv}} \rangle_{\delta} \approx 2 \delta \varepsilon$, confirming our previous statement that a positive vortex merges with the flux as $\beta \nearrow \frac{1}{2}$.
The arguments are similar for $\beta > \frac{1}{2}$, where the roles of $\ell = 0$ and $+1$ are reversed.

Therefore, when $0<\beta <\frac{1}{2}$, the nearest positive vortex is somehow tethered to the flux, and is the cause of the small-$R$, $\beta$-dependent divergence in $\rho$ and $\Delta$ in (\ref{eq:smallR}), whose increase with $\beta$ is due to the approach of the vortex to the flux point. 
The flux at the origin also has positive topological charge, so unlike two regular like-charge vortices which repel, the flux and nearest vortex effectively attract.

\section{Recovering the Aharonov-Bohm Effect}\label{sec:recovering}

We now return to our original objective, to see how the Aharonov-Bohm effect---that is, the phase factor $\exp(2\pi \rmi \alpha)$ for noninteger $\alpha$---is recovered for the integrals of phase $\chi$ around (large) loops enclosing the flux point.
This is related to the total topological charge enclosed by the loop (due to the flux and the vortices) by (\ref{eq:chiint}).
The average phase integral $I(R,\alpha)$ on a circle $C$ of radius $R$ centred at the origin, is simply found from the average azimuthal phase gradient $\overline{\chi_{\theta}}$ (\ref{eq:achi}),
\begin{equation}
   I(R,\alpha) = \frac{1}{2\pi}\oint_C R\langle\partial_{\theta}\chi(R,\alpha)\rangle\rmd\theta = 
   n + R \frac{c}{a},
   \label{eq:chiintI}
\end{equation}
where $c$ and $a$ depend on $\beta$ and $R$, and are given in closed form in (\ref{eq:afull}) and (\ref{eq:cfull}) (as discussed above, $c$ is antisymmetric and $a$ is symmetric in $\beta$, with $\beta$ now defined $-\frac{1}{2}\le \beta \le \frac{1}{2}$). 
$R c/a$ is the total topological charge due to vortices enclosed in the disk of radius $R$, not including the origin.
$n$ is the integer charge due to the flux at the origin.

We have seen that $\rho(R,\alpha) = 0$ for integer and half-integer $\alpha$, so the only contribution to $I(R,\alpha)$ is from the topological charge at the flux. 
When $\alpha = n$, $c = 0$, so $I(R,n) = n$, as we would expect.
When $\alpha = n + \frac{1}{2}$, $c/a = 1/2R$ and so $I(R,n+\frac{1}{2}) = n+\frac{1}{2}$, an average of the integer charges $n$ and $n+1$ as discussed in Section \ref{sec:near}.

\begin{figure}%[!ht]
\centering
\includegraphics[height=120 pt]{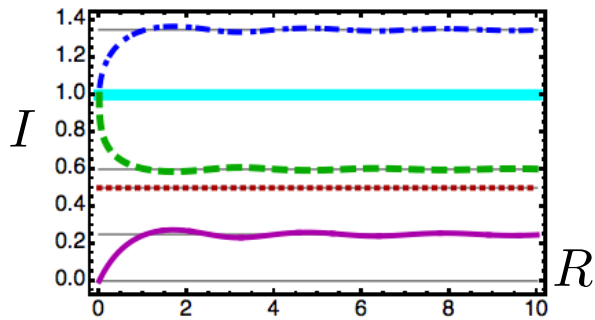}
\caption{
   The integral of the phase gradient over a centred circle of radius $R$ divided by $2\pi$, for various choices of flux $\alpha$: $0.25$, $0.5$, $0.6$, $1$, $1.35$ (solid, dotted, dashed, thick and dot-dashed lines respectively).
   \label{fig:chiint}}
\end{figure}

We would expect that for any flux strength $\alpha$, $I(R,\alpha)$ in (\ref{eq:chiintI}) should behave like the corresponding integral for a deterministic scattered wave: it should tend to $n$ as the circuit's radius $R \to 0$, but give the full AB phase $\alpha$ when $R \gg 1$. 
For small $R,$ the sums of $c$ and $a$ are dominated by the lowest order terms, and approximating from the explicit forms for $\beta>0$, $R c /a \approx \mathrm{constant} \times R^{2-4\beta}$. 
Therefore as $R \to 0$ this term vanishes and indeed $\lim_{R\to 0} I(R,\alpha) = n$.
When $R \gg 1$, from asymptotic expressions of (\ref{eq:afull}), (\ref{eq:cfull}), which we omit, we find $c/a \sim \beta/R$, so for large radius $R$, $I(R,\alpha) \sim \alpha$, indeed recovering the general AB effect.
Of course, this fractional quantity is only a statistical average -- the phase integral around $C$ for any particular sample function must be an integer multiple of $2\pi$.
Plots of $I(R,\alpha)$ against $R$ for different choices of $\alpha$ are shown in Figure \ref{fig:chiint}.

$I(R,\alpha)$ is the average total topological charge inside radius $R$; over the whole plane except the origin, the average sum of all the vortex signs is therefore $2\pi \int_0^{\infty} R \rho(R,\beta)\rmd R = \beta$, the fractional part of the flux strength. 
This may be considered a fractional form of the topological charge screening between vortices, well-studied in the isotropic random ensembles \cite{bd:isotropic,fw:critical,foltin:signed,mrd:correlations}.
Since in the isotropic model the total $\rho$ must be zero, the (suitably normalized) integral of $g_s(R)$ over the plane must give $-1$ to compensate the vortex at $R = 0$, analogous to Coulombic screening in ionic fluids \cite{sl:ions}. 
The second moment of this integral gives a measure of the squared screening length; this in fact diverges for the isotropic random wave model \cite{mrd:correlations} (due to the slow decay of Bessel correlations); similarly, since $R c/a \sim \beta + O(R^{-1})$, the fractional screening length due to the flux also diverges.

\begin{figure}%[!ht]
\centering
\includegraphics[height=120 pt]{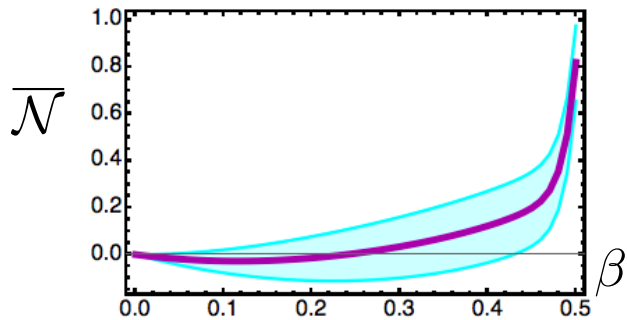}
\caption{
   Average vortex excess $\overline{\mathcal{N}}(\beta)$.
   When $R \gtrsim \pi/2$, the excess oscillates around the mean, with error due to this oscillation as in the shaded area.
   \label{fig:excess}
   }
\end{figure}

The total number of vortices in the disk of radius $R$ centred at but excluding the flux point differs from the expected random wave average of $\pi R^2 \Delta_{\mathrm{irw}}$ by an excess 
\begin{equation}
   \mathcal{N}(R,\beta) = \lim_{\delta \to 0} \int_{\delta}^R 2\pi R'(\Delta(R',\beta) - \Delta_{\mathrm{irw}})\rmd R'.
   \label{eq:excess}
\end{equation}
The approximation (\ref{eq:largeR}) suggests that for $R \gtrsim \pi/2$, the effect of further increasing $R$ merely causes $\mathcal{N}$ to oscillate about a steady mean $\overline{\mathcal{N}}(\beta)$, i.e.~$\mathcal{N}(R,\beta) \sim \overline{\mathcal{N}}(\beta) + \sin(\beta \pi) \sin(2R)/2\pi$, represented in Figure \ref{fig:excess}.
$\overline{\mathcal{N}}(\beta)$ is small for $\beta \lesssim \frac{1}{2}$, then increasing close to $1$, as the nearby vortex migrates towards the flux line. 
Thus  $\overline{\mathcal{N}}(\beta)$ may be interpreted as the statistical confidence of the presence of the additional vortex. 
From (\ref{eq:largeR}) we see that the fluctuations in $\Delta(R)$ are annular rings of width $\pi/2$, with alternating  excesses and deficits, the extra, nearest vortex only becoming statistically visible once it is within $\pi/2$ of the flux, and $\overline{\mathcal{N}}(\beta)$ is well approximated by integrating only up to $R \approx \pi/2$.

\section{Conclusion}

We have investigated a random model of the Aharonov-Bohm effect solving Schr\"odinger's equation in the presence of a flux point.
As with the standard AB wavefunction (\ref{eq:psiAB}), random AB waves have an unobservable integer topological charge $n$ at the flux point, and around an asymptotically large circuit around the flux, the total phase change is $2\pi \alpha = 2\pi(n + \beta)$.
The observable, fractional part of this phase change is due to the random arrangement of $\pm 1$ phase vortices (complex zeros), whose distribution is subtly shifted from the more familiar distribution in isotropic random waves, especially close to the flux point which tends to attract a vortex of $\mathrm{sign}\beta$, merging with it as $\beta$ approaches $\pm\frac{1}{2}$.
The justification of same-sign vortex attraction was supported both using the statistical vortex densities (Section \ref{sec:statistics}) and a small-$R$ approximation (Section \ref{sec:near}).

The flux-adapted random wave model described here falls into the menagerie of random function models which map onto chaotic wave systems, such as boundary-adapted models \cite{berry:statistics,bi:nodal,wheeler:curved}, as well as the chaotic analytic function \cite{hannay:chaotic}, which is a model for wavefunctions in a uniform magnetic field, whose (complex analytic) vortices only have one sign.
In our case, our system is a model for Aharonov-Bohm billiards, and we can compare the results of our analysis with the preliminary results of \cite{br:dislocations}, by looking at the vortex count in a chaotic AB billiard as a function of energy.
As we discussed in the previous section, for sufficiently large distances (i.e.~eigenfunctions above the lowest energies), the overall effect on the vortex count is to add no more than one (with a position-dependent fluctuation also of constant magnitude).
This is consistent with the vortex-count analogue of the Weyl series \cite{ww:acoustics,stoeckmann}, as a flux point acts like the constant (curvature dependent) term of constant order, correcting the first term (area $\times$ density) and second term (perimeter-dependent, explored in boundary-adapted models).
From a wave chaos viewpoint, therefore, the primary effect of a noninteger flux is to break time-reversal symmetry, but apart from a wavelength or two from the flux line, the random wave has statistical euclidean symmetry.

The magnitude of the effect of the flux, manifested through the densities $\rho$ and $\Delta$, appears to have no more of an effect on the distribution of vortices than a single vortex itself.
We speculate that the main effect of the flux is, in fact, in attracting the nearby vortex; the rigid long-range structure of the vortex distribution (exemplified by the divergent topological charge screening length) then guarantees the rest of the arrangement at the correct asymptotic orders.
It would therefore be interesting to study how the position of the nearest vortex is related to the phase integral around circles with large $R$ in particular sample functions, which must be $2\pi$ times an integer but average to $2\pi \beta$.
This might illuminate the difference in relative phase shift between the asymptotic forms of $\rho$ and $\Delta$, and the pure isotropic vortex pair correlations $g$ and $g_s$ from (\ref{eq:gsdef}), (\ref{eq:gdef}), which may be related to the fact that the flux of fractional sign $\beta$ attracts, rather than repels, a like-sign vortex.
The hypothesis could be tested further with a calculation of the 2-point number and charge correlation functions in the random AB model, which now depend on the vectorial displacement of two vortices from the flux; if the main role of the flux is to attract the nearby vortex, the vortex pair correlation functions should differ very little from their isotropic counterparts, even close to the flux.
Such a calculation may also reveal subtle effects of the flux beyond the simple densities considered here.

\ack
We are grateful to Michael Berry, John Hannay and Alexander Taylor for discussions. 
This research was funded by a Leverhulme Research Programme Grant.

\appendix

\section{Gaussian random statistics}\label{sec:appa}

The Bessel function summation techniques of Hansen in \cite{hansen:sums} can be used to evaluate the various Bessel function summation formulae (\ref{eq:adef})--(\ref{eq:fdef}).
In the following expressions, it is assumed that $0 < \beta < \frac{1}{2}$ (the sums $c$ and $e$ are clearly antisymmetric with respect to $\beta$, the others are symmetric).
Using the fundamental Bessel sum
\begin{eqnarray}
\fl   A_{\nu}(x) &=& \sum_{j = 1}^{\infty} J_{j+\nu}^2(x) \nonumber \\
\fl    &=&  -\frac{1}{2}J_{\nu}^2(x) + \frac{4^{-\nu}\nu x^{2\nu} \Gamma(2\nu)}{\Gamma^2(1+\nu)\Gamma(1+2\nu)} \phantom{.}_2F_3\left( \begin{array}{c} \nu\quad \nu+\frac{1}{2} \\ 1+\nu \quad 1+\nu \quad 1+2\nu \end{array}; -x^2\right),
   \label{eq:A2F3}
\end{eqnarray}
where the second line features a $_2F_3$ hypergeometric function, we have
\begin{eqnarray}
\fl   a & = & \frac{1}{2}\left( J_{\beta}^2 + J_{1-\beta}^2 + A_{\beta} + A_{1-\beta}\right), \label{eq:afull} \\
\fl   b & = & \frac{1}{4}\left( J_{\beta} J_{-1+\beta} + J_{-\beta} J_{1-\beta} \right), 
\label{eq:bfull} \\
\fl c & = & \frac{\beta}{2R}(A_{\beta} + A_{1-\beta}) + \frac{1}{4} \left(\left(R+\frac{2}{R}\right) J_{1-\beta}^2+(2 \beta-3) J_{2-\beta} J_{1-\beta} + R J_{2-\beta}^2 \right. \nonumber \\ 
\fl   & & \qquad \left.  - R J_{\beta}^2-R J_{\beta+1}^2+(2\beta +1) J_{\beta}  J_{\beta +1} \right) 
\label{eq:cfull} \\
\fl   d & = & \frac{1}{4}(A_{\beta}+A_{1-\beta}) + \left(\frac{\beta}{8}J_{1-\beta}^2 + \frac{(2-\beta)}{8} J_{-1+\beta}^2 - \frac{\beta(1-\beta)}{4R}J_{1-\beta} J_{-\beta} + \frac{(1+\beta)}{8}J_{-\beta}^2 \right. \nonumber \\
\fl   & & \qquad \left. - \frac{(1 - \beta) \beta}{4R} J_{-1+\beta} J_{\beta} + \frac{(1-\beta)}{8} J_{\beta}^2\right), 
\label{eq:dfull} \\
\fl   e & = & \frac{1}{8R}\left( R J_{1-\beta}^2 - R J_{-1+\beta}^2 + 2 \beta J_{1-\beta} J_{-\beta} + R J_{-\beta}^2 +2 \beta J_{-1+\beta} J_{\beta} - R J_{\beta}^2 \right) \label{eq:efull} \\
\fl f & = & \frac{1}{8R^2}\left\lbrace 2(A_{\beta}+A_{1-\beta})(2\beta^2+R^2)+\left[-4(2+(\beta-4)
\beta)+(2+3\beta)R^2\right]
J_{1-\beta}^2 \right. \nonumber \\
\fl & & \qquad \left. +2R^2J_{2-\beta}^2-3\beta R^2J_{-1+\beta}^2+2\left[4+\beta(3\beta
-5)\right]RJ_{1-\beta}J_{-\beta}+(3\beta-1)
R^2J_{-\beta}^2 \right. \nonumber \\
\fl & & \qquad \left. +2\beta(3\beta-1)RJ_{-1+\beta}J_{\beta}
+\left[4\beta^2+(1-3\beta)R^2\right]J_{\beta}^2  \right\rbrace
\label{eq:ffull}
\end{eqnarray}
where all dependence on $R$ in the Bessel functions has been suppressed.

The various averages, $\langle \partial_{\theta}\chi\rangle, \rho,\Delta \ldots$, are calculated analytically in terms of $a$-$f$.
Any average $\langle F(\bi{X})\rangle$ of a function $F(\bi{X})$ of an $N$-dimensional vector of gaussian random variables $\bi{X}$, whose correlation matrix $\boldsymbol{\Sigma}$ has elements $\Sigma_{ij} = \langle X_i X_j \rangle$, can be expressed
\begin{equation}
   \langle F(\bi{X})\rangle = \frac{1}{(2\pi)^{N/2} \sqrt{\det\boldsymbol{\Sigma}}} \int F(\bi{X}) \exp\left(-\frac{1}{2}\bi{X}\boldsymbol{\Sigma}^{-1}\bi{X}\right)\rmd^N\bi{X}
   \label{eq:grav}
\end{equation}
which can usually be integrated directly provided the form of $F(\bi{X})$ is sufficiently simple.

Now, a derivative of the phase $\chi$ with respect to a variable $i$ (which is $R$ or $\theta$) can be expressed in terms of the real and imaginary parts of $\Psi$, $\partial_i \chi = (\xi \partial_i \eta - \eta \partial_i \xi)/(\xi^2 + \eta^2).$
Therefore the ensemble average of $\partial_{\theta}\chi$ is
\begin{equation}
\fl   \left\langle \partial_{\theta}\chi \right\rangle = \left\langle \frac{\xi \partial_{\theta} \eta - \eta \partial_{\theta} \xi}{\xi^2 + \eta^2} \right\rangle = \int_0^{\infty} \left\langle (\xi \partial_{\theta} \eta - \eta \partial_{\theta} \xi) \exp\left(-u (\xi^2+\eta^2)\right)\right\rangle\rmd u,
   \label{eq:chiavapp}
\end{equation}
where in the last equation, a standard integral representation commutes with the average.
Applying (\ref{eq:grav}) to (\ref{eq:chiavapp}) with $\bi{X} = (\xi,\eta,\partial_{\theta}\xi,\partial_{\theta}\eta)$ (so $\boldsymbol{\Sigma}$ involves only $a$ and $c$), the gaussian integrals are straightforward (incorporating the quadratic form $u(\xi^2 + \eta^2)$ into the gaussian matrix in the exponent), leaving
\begin{equation}
   \left\langle \partial_{\theta}\chi \right\rangle = \int_0^{\infty} \frac{2c}{(1+2 a u)^2}\rmd u = \frac{c}{a},
   \label{eq:chiavapp2}
\end{equation}
confirming the angular equation of (\ref{eq:achi}).
A similar argument applies for $\langle \partial_{R} \chi \rangle$ with $\bi{X} = (\xi,\eta,\partial_{R}\xi,\partial_{R}\eta)$, which vanishes due to the different form of $\boldsymbol{\Sigma}$.

Calculating the density $\Delta$ from (\ref{eq:Deltaxieta}) is rather more complicated, and involves
\begin{eqnarray}
   \Delta & = & \langle \delta(\xi)\delta(\eta) R^{-2}|\partial_R \xi \partial_{\theta} \eta - \partial_{\theta} \xi \partial_R \eta| \rangle \nonumber \\
   & = & -\frac{1}{\pi}\dashint \partial_t \left\langle\delta(\xi)\delta(\eta)\exp\left(\rmi t R^{-2}[\partial_R \xi \partial_{\theta} \eta - \partial_{\theta} \xi \partial_R \eta]\right)\right\rangle \frac{\rmd t}{t}
   \label{eq:Deltaapp}
\end{eqnarray}
where again a standard integral identity has been used \cite{mrd:nodal}, with $\dashint$ denoting a Cauchy principal value integral with a pole at the origin.
We now apply (\ref{eq:grav}) where $\bi{X}$ is 6-dimensional and $\boldsymbol{\Sigma}$ involves all of the averages (\ref{eq:adef})-(\ref{eq:fdef}).
We proceed as before, first integrating the $\delta$-functions, then gaussian integrating over the field derivative variables, whose quadratic form combines the components of $\boldsymbol{\Sigma}^{-1}$ and that of the modulus in (\ref{eq:Deltaapp}).
After all this, there is only the Cauchy principal value integral in $t$,
\begin{equation}
\fl   \Delta = \frac{1}{2\pi^2}\dashint \frac{(\rmi (b c - a e) - [c^2 d - 2 b c e + b^2 f + a (e^2 - d f)] t)}
   {(a + 2 \rmi (b c - a e) t - [c^2 d - 2 b c e + b^2 f + a (e^2 - d f)] t^2)^2}\frac{\rmd t}{t}.
   \label{eq:Deltaapp2}
\end{equation}
The integrand has two complex double poles in the complex $t$ plane as well as the simple pole at the origin.
Writing the principal value integral as the average of the integral of two contours on the real $t$ line, each avoiding the origin in the upper and lower half-plane, we have a residue contribution from each double pole.
The net result of these contour integrals gives (\ref{eq:Deltabcdef}).
The calculation of $\rho$ can be performed directly with this set up without the modulus sign in (\ref{eq:Deltaapp}), and can be shown to give (\ref{eq:rhoabce}), as it must.


\section*{References}

\begin{thebibliography}{99}

\bibitem{l-h:statistical} Longuet-Higgins M S 1957 The statistical analysis of a random, moving surface \PRS \emph{A} \textbf{249} 321--87

\bibitem{ww:acoustics} Wright M and Weaver R eds. 2010 \emph{New directions in linear acoustics and vibration} Cambridge University Press

\bibitem{bd:isotropic} Berry M V and Dennis M R 2000 Phase singularities in isotropic random waves \PRS \emph{A} \textbf{456} 2059--79

\bibitem{l-h:isotropic} Longuet-Higgins M S 1957 Statistical properties of an isotropic random surface \PRS \emph{A} \textbf{250} 157--74

\bibitem{berry:regular} Berry M V 1977 Regular and irregular semiclassical wavefucntions \JPA \textbf{10} 2083--91

\bibitem{stoeckmann} St\"{o}ckmann H-J 2006 \textit{Quantum chaos: an introduction} Cambridge University Press 

\bibitem{bs:autocorrelation} B\"{a}cker A and Schubert R 2002 Autocorrelation function of eigenstates in chaotic and mixed systems \JPA \textbf{38} 539--64

\bibitem{berry:statistics} Berry M V 2002 Statistics of nodal lines and points in chaotic quantum billiards: perimeter corrections, fluctuations, curvature \JPA \textbf{35} 3025--38

\bibitem{ksw:classical} Kuhl U, St\"{o}ckmann H-J and Weaver R 2005 Classical wave experiments on chaotic scattering \JPA \textbf{38} 10433--63

\bibitem{ur:supporting} Urbina J D and Richter K 2003 Supporting random wave models: a quantum mechanical approach \JPA \textbf{36} L485-L502

\bibitem{bohigas:chaos} Bohigas O 1991 Random matrix theories and chaotic dynamics \emph{Les Houches Session LII -- Chaos and Quantum Physics} ed MJ Giannoni, A Voros and J Zinn-Justin (Amsterdam: North-Holland) pp 89--199

\bibitem{AB} Aharonov Y and Bohm D 1959 Significance of electromagnetic potentials in the quantum theory \textit{Phys Rev} \textbf{115} 485--91

\bibitem{op:quantum} Olariu S and Popescu I 1985 The quantum effects of electromagnetic fluxes \RMP \textbf{57} 339--436

\bibitem{aa:AB} Ajiki H and Ando T 1994 Aharonov-Bohm effect in carbon nanotubes \textit{Physica B} \textbf{201} 349--52

\bibitem{bachtold:AB} Bachtold A, Strunk C, Salvetat J-P, Bonard J-M, Forr\'{o} L, Nussbaumer T and Sch\"{o}nenberger C~1999 AharonovÐBohm oscillations in carbon nanotubes \emph{Nature} \textbf{397} 673--5 

\bibitem{peng:AB} Peng H, Lai K, Kong D, Meister S, Chen Y, Qi X-L, Zhang S-C, Shen Z-X, Cui Y 2010 AharonovÐBohm interference in topological insulator nanoribbons \textit{Nature Materials} \textbf{9} 225--9 

\bibitem{br:statistics} Berry M V and Robnik M 1986 Statistics of energy levels without time-reversal symmetry: Aharonov-Bohm chaotic billiards \JPA \textbf{19} 649--68

\bibitem{br:dislocations} Berry M V and Robnik M 1986 Quantum states without time-reversal symmetry: wavefront dislocations in a non-integrable Aharonov-Bohm billiard \JPA \textbf{19} 1365--72

\bibitem{berry:exact} Berry M V 1980 Exact Aharonov-Bohm wave function obtained by applying Dirac's magnetic phase factor \EJP \textbf{1} 240-4

\bibitem{berry:wavefront} Berry M V, Chambers R G, Large M D, Upstill C and Walmsley J C 1980 ``Wavefront dislocations in the Aharonov-Bohm effect and its water wave analogue'' \EJP \textbf{1} 154--62 

\bibitem{mrd:burgers} Dennis M R 2009 On the Burgers vector of a wave dislocation \JOA \textbf{11} 094002

\bibitem{fw:critical} Freund I and Wilkinson M 1998 Critical-point screening in random wave fields \JOSA \emph{A} \textbf{15} 2892--902

\bibitem{sbs:distribution} Saichev A I, Berggren K-F and Sadreev A F 2001 Distribution of nearest distances between nodal points for the Berry function in two dimensions \PR \emph{E} \textbf{64} 036222

\bibitem{mrd:thesis} Dennis M R 2001 Topological singularities in wave fields \emph{PhD Thesis} Bristol University

\bibitem{foltin:signed} Foltin G 2003 Signed zeros of Gaussian vector fields---density, correlation functions and curvature \JPA \textbf{36} 1729--41

\bibitem{mrd:correlations} Dennis M R 2003 Correlations and screening of topological charges in Gaussian random fields \JPA \textbf{36} 6611--28

\bibitem{mrd:nodal} Dennis M R 2007 Nodal densities of planar gaussian random waves \EJP: \emph{Special Topics} \textbf{145} 191--210 

\bibitem{rb:false} Berry M V and Robnik M 1986 False time-reversal violation and energy level statistics: the role of anti-unitary symmetry \JPA \textbf{19} 669--82

\bibitem{hohmann:density} H\"{o}hmann R, Kuhl U, St\"{o}ckmann H-J, Urbina J D and Dennis M R 2009 Density and correlation functions of vortex and saddle points in open billiard systems \PR \emph{E} \textbf{79} 016203 

\bibitem{msg:avoided} Monastra A G, Smilansky U and Gnutzmann S 2003 Avoided intersections of nodal lines \JPA \textbf{36} 1845-53

\bibitem{dlmf} \emph{NIST Digital Library of Mathematical Functions} \url{http://dlmf.nist.gov/}, Release 1.0.13 of 2016-09-16. Olver F W J, Olde Daalhuis A B, Lozier D W, Schneider B I, Boisvert R F, Clark C W, Miller B R, and Saunders B V, eds.

\bibitem{sl:ions} Stillinger F H and Lovett R 1968 General restriction on distribution of ions in electrolytes \JCP \textbf{49} 1991--4 

\bibitem{hansen:sums} Hansen E 1966 On some sums and integrals involving Bessel functions \emph{American Mathematical Monthly} \textbf{73} 143--50

\bibitem{bi:nodal} Berry M V and Ishio H 2002 Nodal densities of Gaussian random waves satisfying mixed boundary conditions \JPA \textbf{35} 5961--72

\bibitem{wheeler:curved} Wheeler C T 2005 Curved boundary corrections to nodal line statistics in chaotic billiards \JPA \textbf{38} 1491--504

\bibitem{hannay:chaotic} Hannay J H 1998 The chaotic analytic function \JPA \textbf{31} L755--61


\end{thebibliography}
\end{document}